\documentclass[3p,times,procedia]{elsarticle}
\usepackage{ecrc}
\usepackage{units}
\usepackage{amssymb}
\usepackage{amsmath}
\usepackage[figuresright]{rotating}
\volume{00}
\firstpage{1}
\journalname{Physics Procedia}
\runauth{}
\jid{phpro}
\jnltitlelogo{Physics Procedia}
\CopyrightLine{2011}{
Elsevier BV. Selection and/or peer-review under responsibility of the organizing committee for TIPP 2011.}
\begin{document}

\begin{frontmatter}
\dochead{TIPP 2011 -  Technology and Instrumentation for Particle Physics 2011}

\title{Development of a TPC for an ILC Detector}
\author[label1,label2]{R. Diener\footnote{Email: ralf.diener@desy.de} on behalf of the LCTPC collaboration}
\address[label1]{Deutsches Elektronen-Synchrotron DESY, Hamburg site, A Research Centre of the Helmholtz Association, Notkestrasse 85, 22607 Hamburg, Germany}
\address[label2]{http://www.lctpc.org}

\begin{abstract}
The ILD concept, one of two proposed detector concepts for the planned 
International Linear Collider (ILC), foresees a Time Projection Chamber 
(TPC) as the main tracking detector. The LCTPC (Linear Collider TPC) collaboration pursues R\&D to develop 
such a TPC based on the best state-of-the-art technology. 

After tests with smaller prototypes, the current efforts focus on studies using a large 
prototype with a diameter of \unit[770]{mm} and a length of \unit[610]{mm}. This prototype can 
accommodate seven read-out modules of a size comparable to the ones 
that would be used in the final TPC. Several prototypes of modules using 
Micromegas or GEM structures as gas amplification were constructed and tested. Besides the 
traditional pad read-out, a pixel read-out based on the 
TimePix chip is studied in these tests with up to 8 TimePix chips on a 
board. The current status and future plans of the R\&D are presented.
\end{abstract}


\end{frontmatter}

\section{Introduction}
\label{sec:introduction}

The International Linear Collider (ILC)\cite{Brau:2007sg,Phinney:2007gp} is a planned linear electron positron collider with a foreseen length of about \unit[31]{km}. Its center of mass energy is tunable from 200 to \unit[500]{GeV}. Further, the options to run the collider at the Z peak and an upgrade to \unit[1]{TeV} are being considered. It will house two detectors which measure in turn using a push-pull concept.

The ILD concept\cite{ref-ildloi}, one of two proposed multipurpose detector concepts for the ILC, foresees a Time Projection Chamber (TPC) as the main tracking detector. Precision physics measurements at the ILC require a very accurate momentum resolution of $\unit[2 \times 10^{-5}]{/GeV/c}$ in the tracking system, which implies a resolution of $\unit[9 \times 10^{-5}]{/GeV/c}$ for the TPC at the planned magnetic field of \unit[3.5]{T}\cite{ref-ildloi}.

The ILD detector is designed to be suited for the particle flow concept, which aims to reconstruct each particle in the best suited sub detector. To do so, a good track separation, very lightweight tracking detectors ---to limit the material in front of the calorimeters--- and a very good pattern recognition\footnote{In $e^+e^- \rightarrow$ ZZ, WW, hZ and HA more than half of the events contain one or more $K^0_S$ or $\Lambda^0$} are necessary.
The ILD TPC as the main tracker fulfills these requirements: It ensures with over 200 space points per track a robust tracking, which allows for an easy pattern recognition and filtering of machine backgrounds. In addition, the TPC is capable of measurements of the specific energy loss as input for particle identification. The material budget is planned to stay below \unit[5]{\%} of a radiation length $X_0$ in the barrel region and below \unit[25]{\%} of $X_0$ at the end caps. The resolution will be $\sigma \leq \unit[100]{\mu m}$ in $r\phi$ at \unit[3.5]{T} and $\leq \unit[500]{\mu m}$ in $rz$ (see \cite{ref-tpcprc2010}).

\section{The ILD TPC}
\label{sec:ildtpc}

The design challenges of the ILD TPC can be divided in four areas. The first one being the mechanical part including the field cage, the end plate and the cathode. Here, the crucial aspects are the layout and the mechanical accuracy to ensure a very homogeneous electric field. In addition, the limited material budget and the electrical stability have to be considered. The second area is the amplification structure at the anode plate. To meet the design goals, Micro Pattern Gas Detectors (MPGD) have been chosen for the gas amplification. They promise to provide the needed electrical and mechanical stability, while allowing for the coverage of large areas with minimal dead space and a minimized ion back flow in the drift volume. The third area comprises the readout system. Currently, traditional pad and novel pixel chip options are being pursued. Here, the compactness and integration of the amplification and the readout electronics in the end plate, as well as cooling and power pulsing are the critical points. The last area is the reconstruction. Here, calibration and alignment have to be considered. In addition, methods to correct for field inhomogeneities have to be developed and implemented to be able to meet the resolution goals.

To pursue the research and development to develop such a high-performance TPC, the LCTPC\cite{ref-lctpc} collaboration was formed, including groups from the Americas, Europe and Asia. Its research plan can be divided in three phases:\\
First, the Demonstration Phase. This has been ongoing for several years and will continue further. It includes the work with small prototypes to prove the feasibility of a MPGD TPC, to understand point resolution and reconstruction and to map out the parameter space. For CMOS-based pixel TPC ideas this includes proof-of-principle tests.\\
The second phase is called the Consolidation phase. This is currently ongoing and includes the design, the construction and the running of the Large Prototype to test manufacturing techniques for MPGD end plates, field cage and electronics, and to collect experience for the construction of the final TPC in a large detector. Also, the goal is to understand and optimize the momentum resolution, track separation and specific energy loss.\\
The third phase is called the Design Phase and should start around 2011/12. Here, the decision will be taken as to which end plate technology will be used for the linear collider TPC. Further, the design for the TPC will be finished leading to a ``conceptual'' engineering design.

\section{The Large TPC Prototype}
\label{sec:lp}

The Large Prototype (LP) TPC\cite{Behnke:2010ze} was built to study the momentum resolution and dE/dx measurement. In addition, the technical challenges are studied, such as how to scale the readout structures developed in tests with small prototypes to a size scale of modules which could be used in a large ILD TPC. This includes designs that allow a large area coverage with a minimum amount of space and the integration of readout electronics including cooling. Further, the goal is to learn how to build a large scale, lightweight and mechanically precise field cage, anode end plate and cathode. 

\subsection{LP Field Cage}
\label{subsec:lpfieldcage}

\begin{figure}[htb!]
\centering
\includegraphics[width=0.36\textwidth]{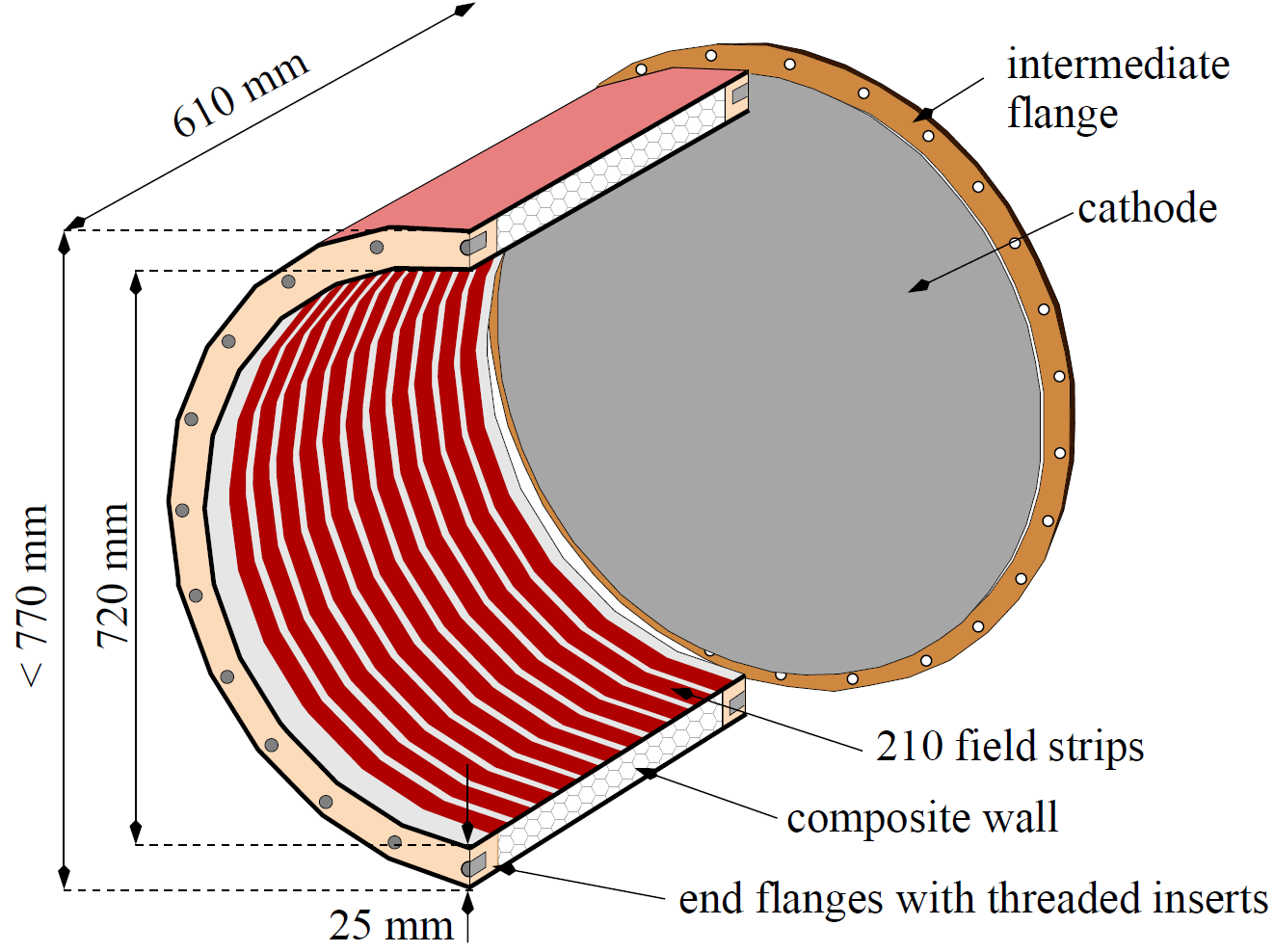}
\includegraphics[width=0.36\textwidth]{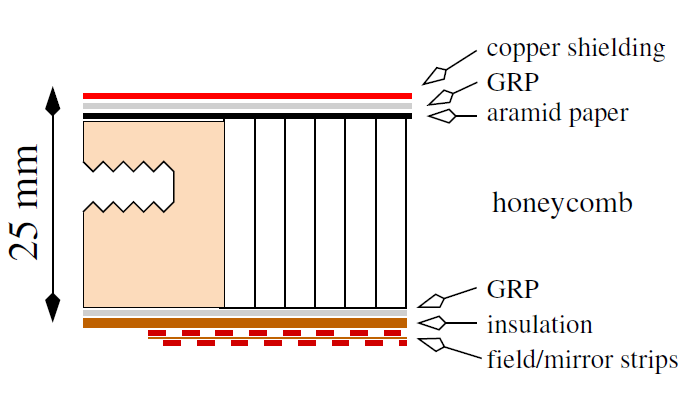}
\includegraphics[width=0.26\textwidth]{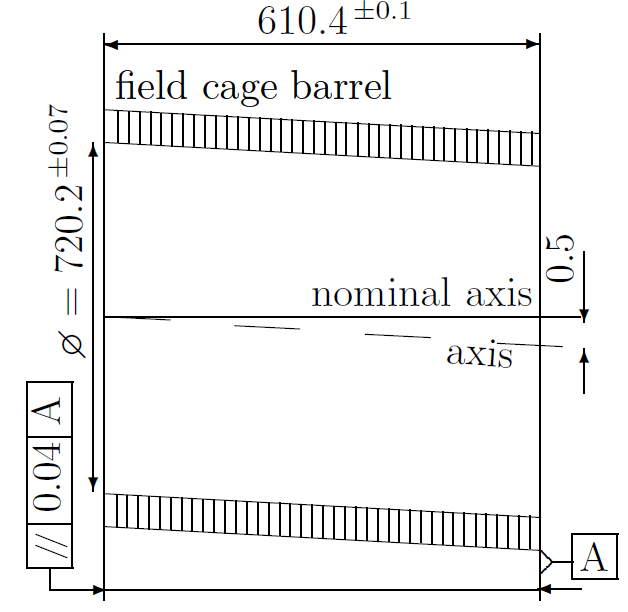}
\caption [Large Prototype]{Left: Schematics of the Large Prototype (LP) TPC. Middle: Cut though the wall of the LP TPC. Right: Sketch of the misalignment of the LP TPC axis.}
\label{fig:lp_sketch}
\end{figure}

A schematic sketch of the LP TPC field cage and cathode is shown in Figure \ref{fig:lp_sketch}. On the left side, its dimensions are shown. It has an inner length of \unit[610]{mm} ---of which in operation \unit[600]{mm} are available as drift length--- and an inner diameter of \unit[720]{mm}. In the middle, a cut through the \unit[25]{mm} thick wall is shown including the end flange at the anode side. It is built of composite materials ---to ensure a lightweight structure and to provide a high mechanical stability--- including additional layers for field shaping and insulation.

Due to a fabrication imperfection, the current LP TPC does not meet the accuracy specifications that are needed to ensure an electric field of a homogeneity of better than $\Delta E / E < 10^{-4}$. The axis of the field cage barrel shows a shearing from the nominal axis of \unit[0.5]{mm} (Figure \ref{fig:lp_sketch}, right), while \unit[0.1]{mm} would be acceptable. The reason for this is an inaccurate mandrel on which the field cage was glued. The mandrel has been worked over and measured and efforts to construct a second field cage are ongoing.

\subsection{Anode End Plate}
\label{subsection:anode}

\begin{figure}[htb!]
\centering
\includegraphics[height=0.25\textwidth]{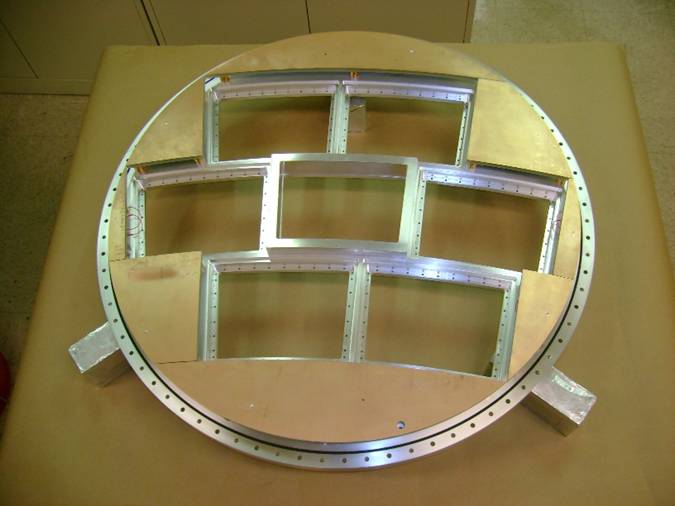}
\hspace{0.05\textwidth}
\includegraphics[height=0.25\textwidth]{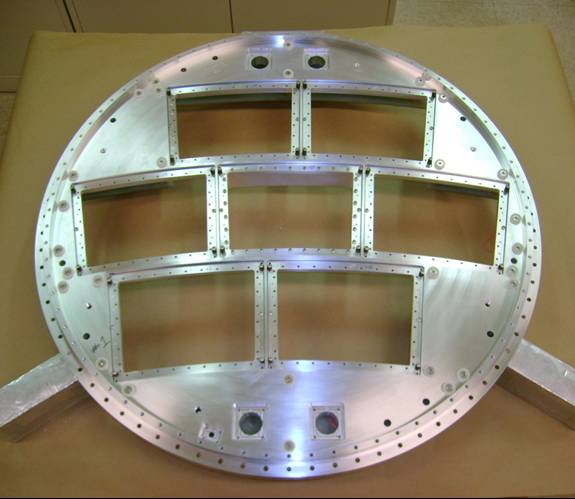}
\caption [Anode End Plate LP1]{First version of LP TPC end plate seen from the inside of the chamber (left side) and from the outside (right side).}
\label{fig:endplatev1}
\end{figure}

The anode side readout end plate was designed to house seven readout modules of a shape and size comparable to one that could be used in the ILD TPC (Figure \ref{fig:endplatev1}). They are arranged in three rows in the end plate, which have a curvature that corresponds to the radius of the outermost ring of readout modules in the ILD TPC end plates.

The R\&D plan for the anode end plate foresees several steps. In the first step, a so-called LP1 end plate of solid aluminum was designed to meet the precision specifications required for the ILD TPC (Figure \ref{fig:endplatev1}). This end plate is currently used during measurements with the LP TPC. Its precision features are accurate to $\sim \unit[30]{\mu m}$. This accuracy is achieved with a 5-step machining process developed at Cornell University. However, this iteration of the end plate does not meet the material limit specified for the ILD TPC; its material budget is 16.9\% X$_0$, which is about two times larger than the goal of 8\%.

\begin{figure}[hbt!]
\centering
\includegraphics[height=0.23\textwidth]{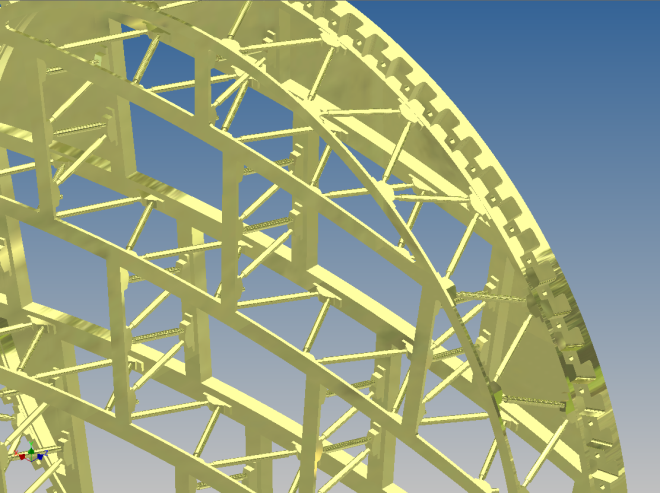}
\hspace{0.05\textwidth}
\includegraphics[height=0.23\textwidth]{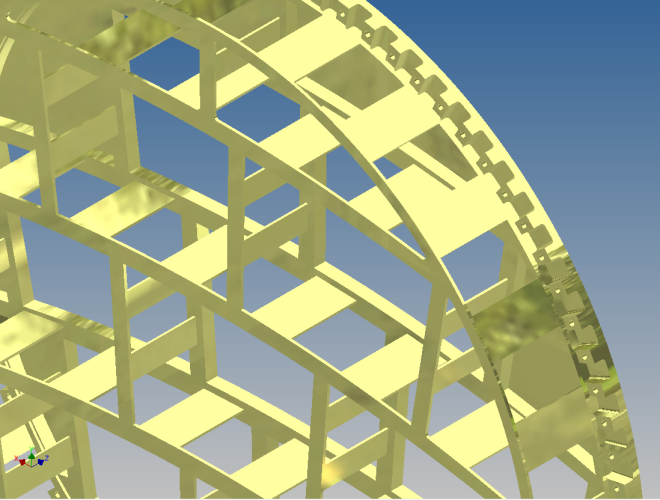}
\caption [Anode End Plate LP1.1]{Designs of a lightened LP1.5 end plate design used in FEA calculations. Left: ``Strut'' space frame design. Right: ``Equivalent plate'' design.}
\label{fig:endplatev1.5}
\end{figure}

The next step in the development, which is currently under way, is to lighten the structure. For this, design results from several studies are being used. These include the experiences with the  current LP1 end plate. Further, Finite Element Analysis (FEA) models of space frame designs for a so-called LP1.5 end plate have been performed (Figure \ref{fig:endplatev1.5}). Two designs have been studied: a ``strut'' space frame design and an ``equivalent plate'' design, which offers the same mechanical accuracy and material budget as the strut design. The FEA calculations for the LP1.5 end plate are backed up by comparisons between simulation and measurements with small test beams that represent one diameter of the LP1 end plate.

The two studied space frame designs have a material budget of 7.5\% X$_0$ and show  in the FEA calculations a deflection of $\unit[23]{\mu m}$ under a load of \unit[100]{N} (\unit[22]{lbs}) in the center module. The construction of the LP1.5 end plate is currently under way following the strut design. Future studies will include construction and measurements using this new LP1.5 end plate.

\section{Readout for the LP TPC}
\label{section:module}

\subsection{Readout Modules}

As mentioned above, the LP TPC end plate can house up to seven readout modules. These are designed to be of similar size and shape to modules as foreseen for the final ILD TPC. Their dimensions are about $23 \times \unit[17]{cm^2}$. 

\begin{figure}[htb!!]
\centering
\includegraphics[height=0.38\textwidth]{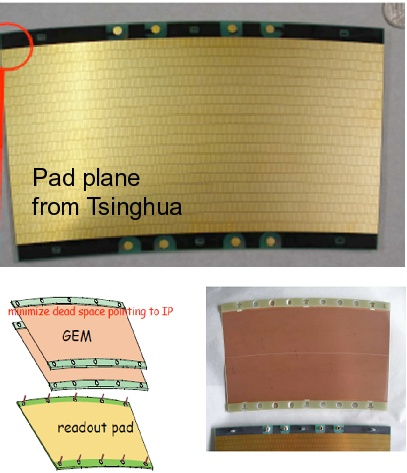}
\hspace{0.02\textwidth}
\includegraphics[height=0.38\textwidth]{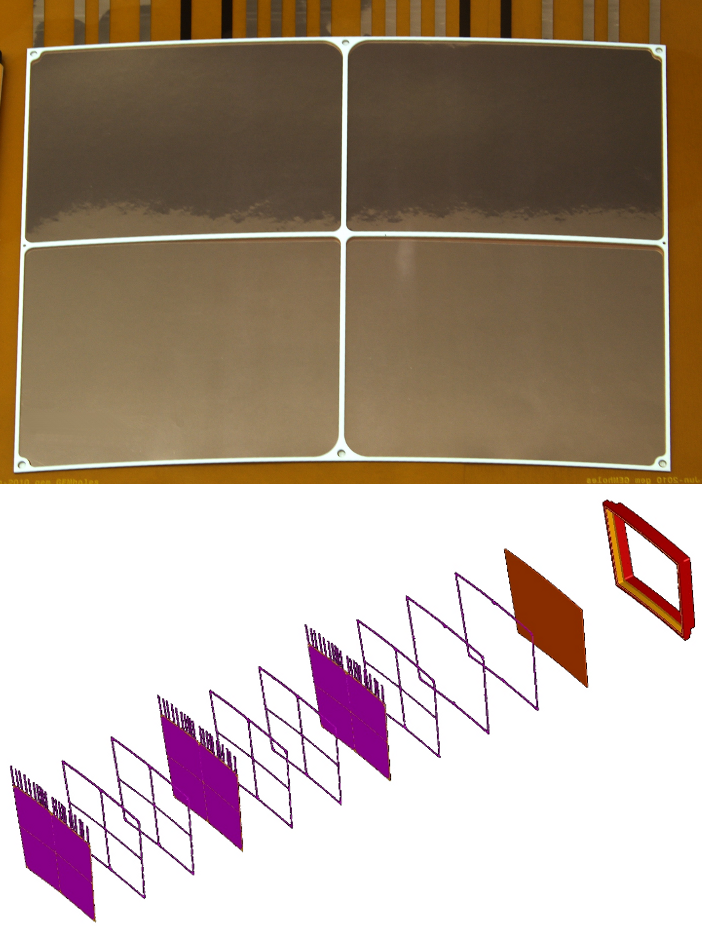}
\hspace{0.02\textwidth}
\includegraphics[height=0.40\textwidth]{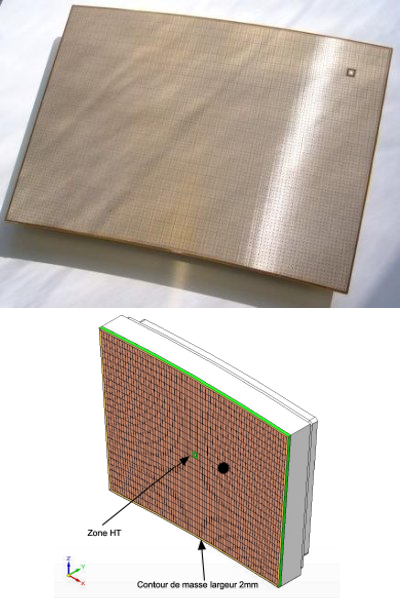}
\caption [LP Modules with Pad readout]{LP readout modules with pad readout. Left: Pad plane, design and GEM foil of a module using double GEM amplification with foils stretched between two bars at the top and bottom of the module. Middle: GEM foil mounted on a ceramic grid and design sketch of a module using triple GEM amplification. Right: Photo and sketch of a module using Micromegas amplification and a resistive layer on the pad plane for charge spreading.}
\label{fig:PadModules}
\end{figure}

Several implementations of such modules using different combinations of gas amplification (GEMs and Micromegas) and readout (pad and pixel chip based) are available. Figure \ref{fig:PadModules} shows three implementations of modules based on pad readout, using pads of sizes ranging from $(1-3) \times \unit[(4-6)]{mm^2}$: two modules using GEM amplification ---one using bars on the top and bottom to stretch the GEM foils between and the other using ceramic grids glued on the GEMs for mounting--- and one using Micromegas and a resistive coating on the pad plane for protection and charge spreading.

As well as the traditional pad readout, a novel readout technique based on the TimePix chip is studied. The pixels of this chip have a size of $55 \times \unit[55]{\mu m^2}$. This readout promises the ``ultimate'' resolution, which is only  limited by the transverse diffusion of drift electrons and not by the geometric influence of the readout pads. One of the main challenges here is to find a working point with a sufficiently large single electron efficiency.

\begin{figure}[hbt!]
\centering
\includegraphics[height=0.23\textwidth]{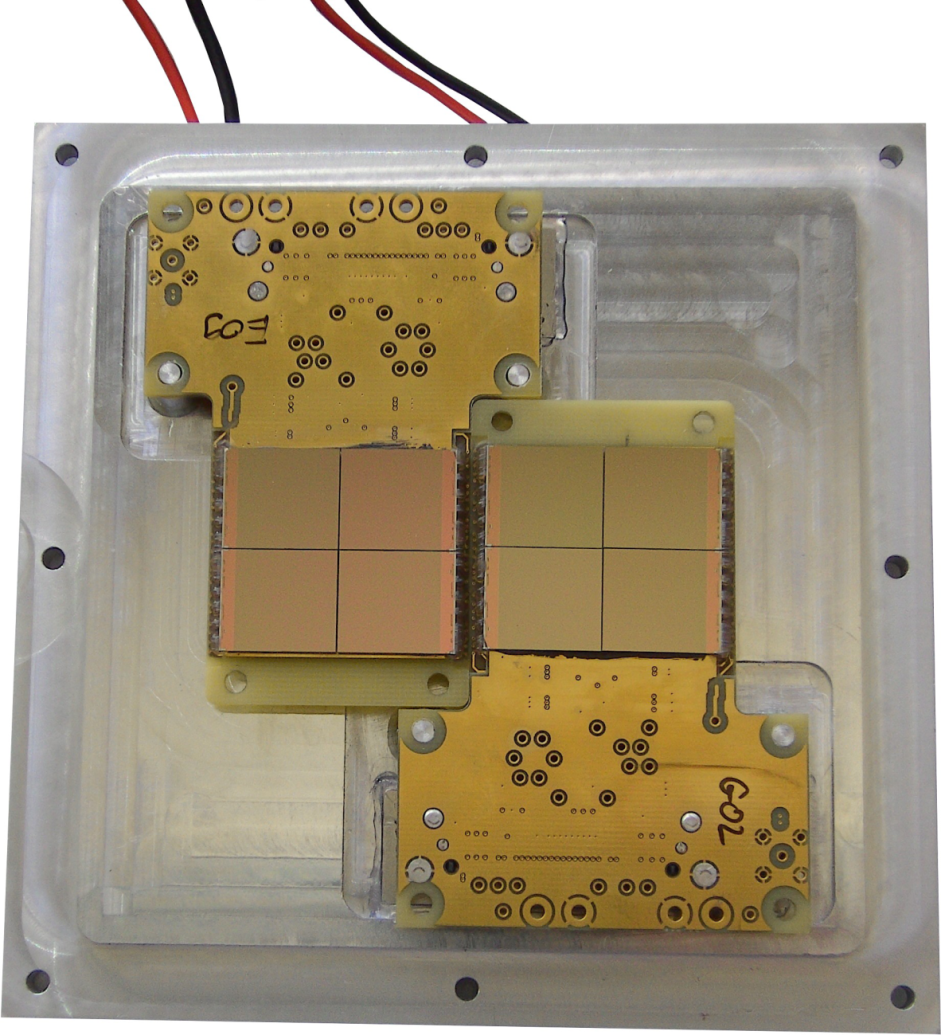}
\hspace{0.05\textwidth}
\includegraphics[height=0.21\textwidth]{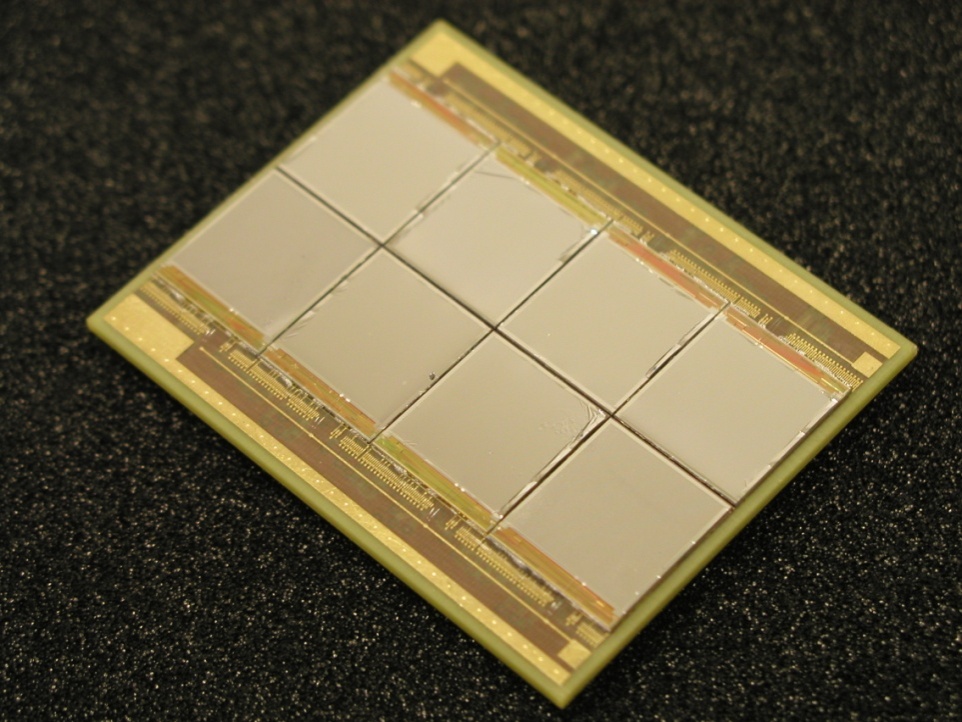}
\caption [TimePix chip setups]{TimePix eight chip boards used in LP modules. Left: Two quad boards used with triple GEM amplification. Right: ``Octopuce'' board with eight Ingrids (integrated Micromegas mesh and Timepix chip).}
\label{fig:timepix}
\end{figure}

Two modules have been tested at the LP: one uses triple GEM amplification on top of two quad chip boards (Figure \ref{fig:timepix}), the other uses eight so-called Ingrids ---a Micromegas like grid integrated directly onto the Timepix chip using CMOS post-processing techniques. While with GEM amplification a signal from a primary electron spreads due to diffusion between the GEMs over 40 to 100 pixels, the Ingrids measure single electrons hitting each one pixel.

\subsection{Readout Electronics}

Currently in use for the readout of the modules is the modified ALTRO electronics from the ALICE\cite{ref-alice} TPC including a new programmable charge sensitive preamplifier, the PCA16 chip, for GEM modules and the AFTER electronics from the T2K\cite{Abe2011106} experiment for the Micromegas modules. For the near future, it is planned to continue the integration work with the PCA16 and ALTRO respectively SALTRO16 chips or the AFTER electronics. None of these are the final ILD electronics since they do not offer sufficient packaging or protection and have too high a power consumption and not enough memory depth. Therefore, it is foreseen to start the design work on a future GdSP (Gaseous detector Signal Processing) chip using synergies between the ILD TPC and the SLHC muon chambers R\&D.

Besides the work on the electronics itself, R\&D for cooling solutions has started. Its goal is to minimize the power consumption and develop a scalable cooling solution to cool the electronics components, to keep the gas temperature constant and to avoid heat radiation to surrounding detectors. One measure is to implement power pulsing, which means to switch off electronics between bunch trains. This can reduce the running time of the readout down to about 1\% of the duty cycle. The second measure is actively cooling the components using 2 phase CO$_2$ or sub-atmospheric water cooling. To study active cooling solutions, a test board with equivalent heat production has been constructed and first tests have started.

\section{Large Prototype Testbeam Setup at DESY}
\label{section:t24}

\begin{figure}[hbt!]
\centering
\includegraphics[width=0.40\textwidth]{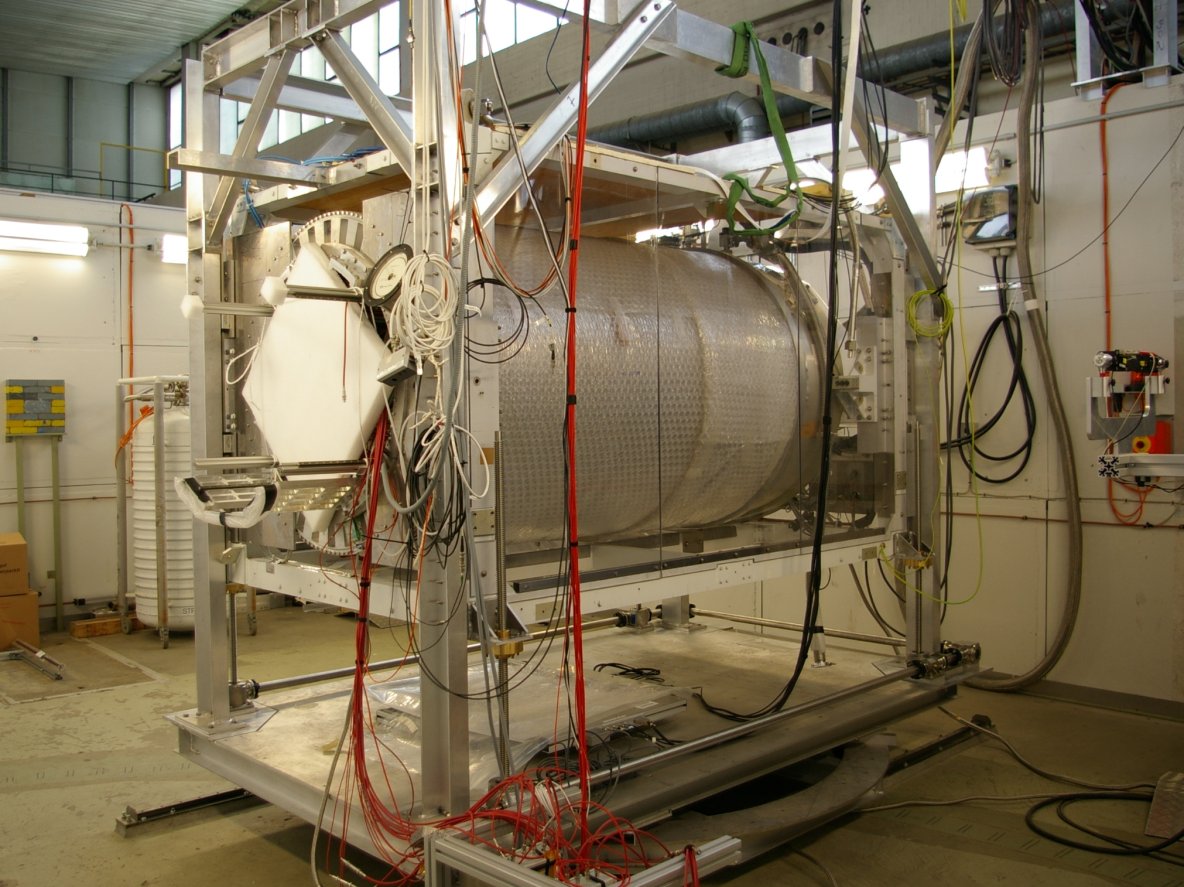}
\caption [T24 test beam area]{T24 test beam area at DESY: movable lifting stage with PCMAG \unit[1]{T} magnet.}
\label{fig:t24}
\end{figure}

Measurements with the LP TPC have been performed in the test beam area T24/1 at DESY, where electron and positron beams from 1 to \unit[6]{GeV} are available. This area hosts a sophisticated infrastructure consisting of equipment that is necessary in order to operate a gaseous detector according to the scientific
program of the LCTPC collaboration. The setup (Figure \ref{fig:t24}) comprises a \unit[1]{T} magnet called PCMAG, which is currently run using persistent current and a LHe reservoir, that is filled using LHe dewars. In the second half of 2011, the magnet is modified to run with cryo coolers and a closed cooling circle. The PCMAG, including a cosmic trigger setup using scintillator slabs, is mounted on a movable lifting stage. During the absence of the magnet due to its modification, the lifting stage will be finished and its control and safety systems completed.

Further, the setup includes a high voltage and a gas system including a slow control setup. In addition, a photo electron calibration system for the LP TPC is available. This consists of a pattern on the cathode plate, from which electrons are liberated when being illuminated by pulses from a UV laser. For the future, the inclusion of silicon detector layers from the former Zeus vertex detector is planned to be capable of doing reference measurements on both sides of the TPC.

\section{Ion Back Flow}
\label{section:ionbackflow}

A critical point for the operation of the ILD TPC could be the back flow of ions from the amplification area into the drift volume. Despite the intrinsic suppression of ion back drift of MPGD detectors, a non negligible amount of ions can still escape the amplification stage. This leads to the effect that after each bunch train of the ILC, a disk of positively charged ions slowly drifts through the sensitive volume to the cathode. Based on the time structure of the ILC bunch trains, this can lead to up to three of these ion disks drifting through the ILD TPC at a point in time. These disks catch electrons from the primary ionization by particles, but more importantly they could lead to field inhomogeneities that deteriorate the measurement precision beyond acceptable values.

Several groups of the LCTPC collaboration are working on calculations and simulations of this effect and to minimize ion back drift. Further, work on hardware setups is ongoing to first, measure the amount of the effect and second, to study possibilities of gating between two bunch trains to stop the ion disk at just upstream of the amplification area. Here ideas are being pursued to use traditional wire gating, but also to use GEMs or meshes for gating.

\section{Software}
\label{section:software}

To enable the R\&D groups to perform detailed analysis and simulations that are comparable and to use synergies between the different groups, a common software package called MarlinTPC\cite{ref-marlintpc_hp} is being developed. It is based on the common ILC software frameworks LCIO\cite{ref-lcio_hp} (data format, persistency), Marlin\cite{ref-marlin_hp} (processing chains), GEAR\cite{ref-gear_hp} (geometry description) and LCCD\cite{ref-lccd_hp} (conditions data handling). 

The MarlinTPC package consists of a detailed simulation down to the single electron level as well as the reconstruction and analysis parts. The current efforts focus on completion and optimization of the basic reconstruction. This includes efforts to supply implementations of hit reconstruction for all readout module types and the implementation/integration of different track finding and fitting packages. The next steps will focus on corrections and simulation. Here, calibration and correction methods for inhomogeneous fields and mechanical alignment have to be included and optimized. Further, the included detailed simulation package for TPC prototypes will be revised and extended. One possible solution here is the use of Garfield++\cite{ref-garfieldpp} in collaboration with the RD51 collaboration\cite{ref-rd51_hp}.

For ILD physics simulations, a realistic model of the ILD TPC is available for Geant4 simulations including digitization, which is based on R\&D results. Currently the complete tracking code for the ILD reconstruction is under revision.

\section{Towards the ILD TPC}
\label{section:towardsild}

\begin{figure}[hbt!]
\centering
\includegraphics[width=0.58\textwidth]{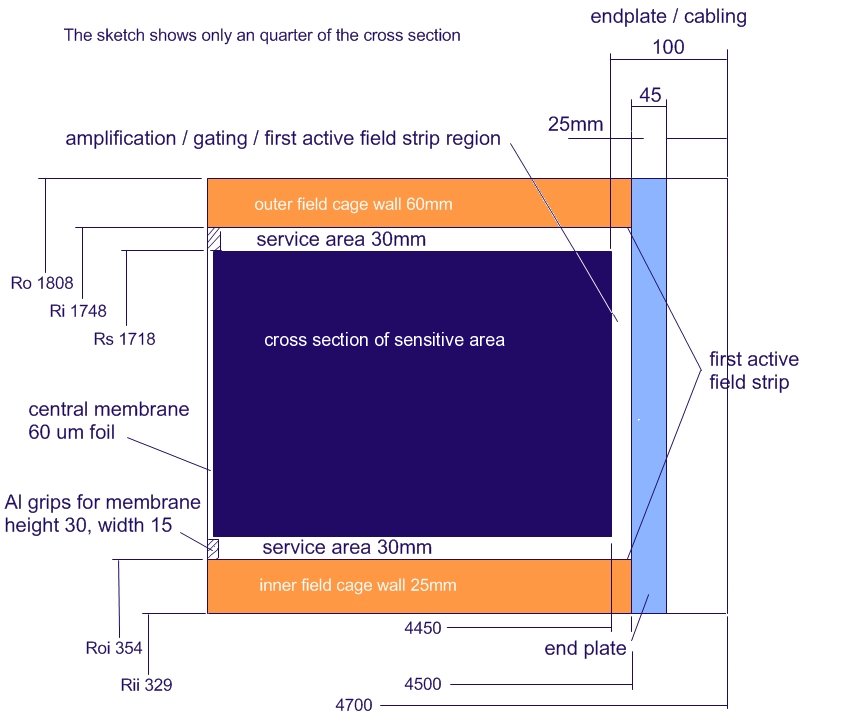}
\caption [ILD TPC dimensions]{Quadrant of the ILD TPC showing its dimensions and structure including space for cabling and supply lines.}
\label{fig:ildtpc}
\end{figure}

The work for the third phase of the R\&D of the LCTPC collaboration, which includes the design of the final ILD TPC, has started. For the field cage design, contact to an external company has been established to perform calculations of the mechanical properties of the composite materials at a large scale. This field cage has to stay lightweight while providing a high mechanical precision. Also, the integration of the ILD TPC in the ILD detector is ongoing in consideration of the surrounding detectors, including mounting structures and spaces for supply lines and cables. Figure \ref{fig:ildtpc} shows the layout and dimensions of the ILD TPC.

Also, work on a cathode design has started. While the cathode of the LP TPC consists currently of a copper coated GRP (glass-reinforced plastic) plate, for the final cathode, designs based on composite materials or coated foils are under consideration. Tensile tests in one direction with different kinds of foils have started and a device to perform two-dimensional tests is being developed. Also, the mounting structure for the cathode foil is being designed.

\begin{figure}[hbt!]
\centering
\includegraphics[height=0.23\textwidth]{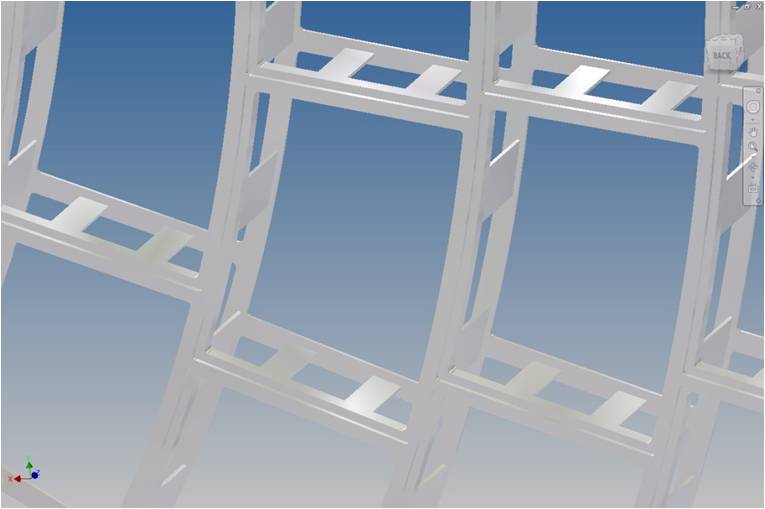}
\hspace{0.01\textwidth}
\includegraphics[height=0.23\textwidth]{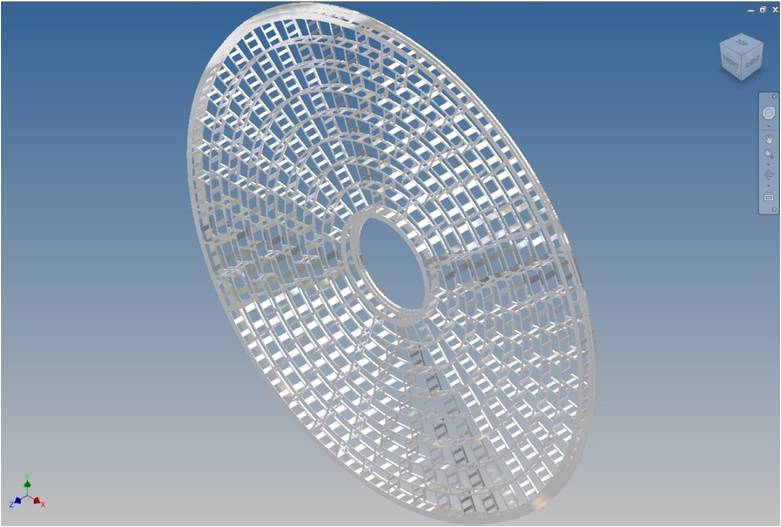}
\caption [ILD End Plate]{FEA Model of the ``equivalent-plate'' design of the ILD end plate.}
\label{fig:ildendplate}
\end{figure}

The ILD end plate design will be a space-frame design similar to one of the LP1.5 end plate (see Section \ref{subsection:anode}). It is shown in Figure \ref{fig:ildendplate} as the solid model used for FEA calculations. Its full thickness is \unit[100]{mm} with a radius of \unit[1.8]{m} and a mass of \unit[136]{kg}. The end plate has a material thickness of \unit[1.34]{g/cm$^2$}, which corresponds to  6\% of a radiation length X$_0$. Shown here is the so-called ``equivalent plate'' design, where the separating members are thin plates. This design has the rigidity and material equivalent to a strut design, which will be used for the LP1.5 end plate. The future ILD design can be either the ``strut'' or the ``equivalent plate'' design.

\section{Summary and Outlook}
\label{section:summary}

For the development of a TPC as the main tracker at the ILD detector for the future ILC e$^+$e$^-$ collider the LCTPC collaboration has formed. In this collaboration, the Large Prototype TPC has been built including a measurement setup, that is used at the DESY electron/positron test beam area by R\&D groups worldwide. Several types of readout modules have been tested and further tests are planned. These plans include tests with seven Micromegas modules, tests with the SALTRO16 electronics and the move of the setup to a hadron test beam. A second iteration of the field cage, including precision improvements and a new end plate, is foreseen for 2012.

Besides the work with the Large Prototype, efforts towards the design of the final TPC at the ILD detector, including R\&D and design activities on the mechanics, readout electronics, cooling solutions and their integration, have begun. The open question of backgrounds, ion back flow and gating are identified and possible solutions are being studied. For this, also detailed studies with small prototypes are being performed. Other open issues in hardware and software development are identified and being addressed.

\section{Acknowledgments}

This work was supported in part by the U.S. National Science Foundation, by the  Commission of the European Communities under the 6th Framework Program ``Structuring the European Research Area'' under contract RII3-026126 and by the Swedish Council (VR).

\bibliographystyle{elsarticle-num}
\bibliography{tpc2010LOI19}

\end{document}